\documentclass[showpacs,showkeys,preprintnumbers,amsmath, amssymb,latexsym,floatfix,superscriptaddress,onecolumn,prd]{revtex4-1}

\usepackage{graphicx}
\usepackage{dcolumn}
\usepackage{bm}

\usepackage{color,xcolor}
\newcommand{\us}{\textcolor{black}}

\makeatletter    

\def\meqalign#1{\null\,\vcenter{\openup\jot\m@th \let\\=\crcr 
  \ialign{\strut\hfil$\displaystyle{##}$&&$\displaystyle{{}##}$\hfil
      \crcr#1\crcr}}\,}

\makeatother 	
\begin{document}

\title{Charged cosmological black hole}

 \author{Rahim Moradi}\email{Rahim.Moradi@icranet.org}
 \affiliation{\ Dipartimento di Fisica and ICRA, Sapienza Universit\`a di Roma, P.le Aldo Moro 5, I--00185 Rome, Italy.}
\affiliation{\ ICRANet, Piazza della Repubblica 10, I--65122 Pescara, Italy.}
 \author{Cl\'{e}ment~Stahl}\email{clement.stahl@pucv.cl}
\affiliation{Instituto de Fisica, Pontificia Universidad Catolica de Valparaiso, Casilla 4950, Valparaiso, Chile} 
\author{Javad~T.~Firouzjaee}\email{j.taghizadeh.f@ipm.ir}
\affiliation{\ School of Astronomy, Institute for Research in Fundamental Sciences (IPM)
P.O. Box 19395-5531, Tehran, Iran.}
\author{She-Sheng~Xue}\email{xue@icra.it}
\affiliation{\ ICRANet, Piazza della Repubblica 10, I--65122 Pescara, Italy.}

\date{\today}


\begin{abstract}

The cosmological black holes are black holes living not in an asymptotically flat universe but in an expanding spacetime. They have a rich dynamics in particular for their mass and horizon. In this article we perform a natural step in investigating this new type of black hole: we consider the possibility of a \textit{charged} cosmological black hole. We derive the general equations of motion governing its dynamics and report a new analytic solution for the special case of the charged Lemaître-Tolman-Bondi equations of motion that describe a charged cosmological black hole. We then study various relevant quantities for the characterization of the black hole such as the C-function, the effect of the charge on the black hole flux and the nature of the singularity.  We also perform numerical investigations to strengthen our results. Finally we challenge a model of gamma ray burst within our framework.

\end{abstract}

\maketitle

\section{Introduction}
\label{sec:intro}
\par
The theoretical tools to model black holes (BH) are nowadays well known through uniqueness theorems \cite{Chrusciel:2012jk} that state that any collapsing structure or black hole merger will reach after enough time a Kerr solution which was described in the 60's \cite{Kerr:1963ud}. Having a Kerr solution, a lot of working power has been dedicated in order to propose generalizations of those astrophysical black holes to higher spacetime dimension \cite{Klebanov:1997kv}, with new matter fields included and/or with different ways of modeling gravitational processes than Einstein general relativity (GR) \cite{delaCruzDombriz:2012xy}. Much less has been done, first to investigate the impact the asymptotic state of the black hole on its dynamics. Usually, black holes approach asymptotically a Minkowski flat space time but it is legitimate to question whether an asymptotically expanding universe would change the status of the black hole. And second many unknowns remains regarding collapsing structures and the process of formation of a black hole. Constructing such cosmological collapsing structures is not only useful to explore non-linear effect of GR but also to explore quasi-local features of that structure such as masses and horizons \cite{Firouzjaee:2008gs}, black hole thermodynamics and Hawking radiation \cite{Giddings:2012bm,Firouzjaee:2011hi,Firouzjaee:2014zfa,Firouzjaee:2015bqa} or the validity of the weak field approximation \cite{Mood:2013mwa}.
\par
It is today a textbook statement that our universe is expanding that is the background of any physical process is the Friedmann-Lemaitre-Robertson-Walker (FLRW) metric. Some research groups are starting to implement this idea into various astrophysical and cosmological setups when the physical processes under scrutiny are sensitive to this subtle difference. As a matter of illustration,  the production of gravitational waves are usually defined with respect to an asymptotic Minkowski spacetime but have a totally different interpretation and pattern when one considers a de Sitter background \cite{Ashtekar:2015ooa,Ashtekar:2014zfa,Ashtekar:2015lxa,Ashtekar:2015lla}. Here we adopt this point of view for black holes and therefore consider a cosmological black hole (CBH) defined as a structure collapsing within an expanding universe after the radiation era \cite{Moradi:2015caa}. The main difference from an astrophysical black hole is that while ABH are static or asymptotically stationary, CBH have a dynamical horizon and a dynamical mass function. The lensing properties (the deviation angle and the time delay) of the CBH were shown to be different than the ABH thus making a testable prediction for this modeling of black hole \cite{Mood:2013uba} ; with the data available no significant difference was however reported.
\par
The second motivation to consider CBH is that they also describe the forming process of a collapsing structure which, from numerical simulations, are known to have a non-static horizon \cite{Dreyer:2002mx,Ashtekar:2004cn} before asymptotically reaching the Kerr solution from the uniqueness theorems. When one models a collapsing structure one of the technicalities appearing is that global concepts of black holes such as event horizons cannot be defined in the non-stationary and asymptotically FLRW model. The need for a local definition of black holes and their horizons has led to concepts such as Hayward's trapping horizon \cite{Hayward:1993wb}, isolated horizon \cite{Ashtekar:2000sz}, Ashtekar and Krishnan's dynamical horizon \cite{Ashtekar:2002ag,Ashtekar:2003hk}, and Booth and Fairhurst's slowly evolving horizon \cite{Booth:2003ji}. The explicit black hole solution that we will present in this work can help to explore the differences and relations between those different horizons.
\par
Several attempts to model collapsing structures into an asymptotically expanding background exist in the literature \cite{Einstein:1945id,Valkenburg:2011tm,Gao:2011tq,Abdalla:2013ara,Bibi:2017urt}, for instance the McVittie solution \cite{McVittie:1933zz} is still an active field of research \cite{Nolan:1999kk,Faraoni:2014nba} even if some concerns have been raised \cite{Kaloper:2010ec,Lake:2011ni}. In the McVittie model, the way the dynamic collapsing structure is tailored implies that the matter field in this solution of GR is restricted. Conversely, the Lemaitre-Tolman-Bondi (LTB) solution \cite{Lemaitre:1933gd,Tolman:1934za,Bondi:1947fta}, describes an isotropic but inhomogeneous spacetime filled with a dust fluid. Because of the freedom of the free functions of the metric, the model can describe a collapsing structure without the restriction on the matter field present in the McVittie solutions. It is furthermore possible to (asymptotically) recover the FLRW metric from the LTB one for a suitable choice of the free functions of the LTB metric. The LTB metric has been applied to various physical systems ranging from modeling radial inhomogeneities \cite{Bolejko:2011jc} or fractal pattern \cite{Stahl:2016vcl} on cosmological scales, to investigate singularity theorems or nucleus in nuclear physics \cite{Pleb}, it is shown in \cite{Firouzjaee:2008gs,Firouzjaee:2011dn} that the LTB metric admits a cosmological black hole solution.
\par
In this article, we complete the literature on cosmological black holes by making a natural step forward: in the same way the Schwarzschild solution generalizes to the Reissner–Nordstr$\ddot{\text{o}}$m (RN) solution by considering the coupled Einstein-Maxwell system, we will generalize the neutral CBH to its charged counterpart: the charged cosmological black hole (CCBH). A similar step has been performed for the McVittie solution \cite{Gao:2004cr}. 
\par \us{A traditional belief in many astrophysical systems is that the electric charge can be negleted:} in the classical works of Wald \cite{Wald:1974np} and Blandford-Znjek \cite{Blandford:1977ds} it is believed that black holes with large charge-mass ratio don't exist in nature. Wald \cite{Wald:1974np} has shown that the charge-mass ratio for a Kerr black hole rotating in the small uniform magnetic field of a galaxy ($ 10^{-4} - 10^{-5} \text{ Gauss}$) is $ \simeq 10^{-24}$. However, if a highly magnetized plasma accretes onto the black hole, the charge-to-mass ratio can be much larger. In particular, in the merging of a binary system of neutron stars, it is expected at the final steps of a gravitational collapse to a black hole to obtain electromagnetic fields larger than the critical value for vacuum polarization \cite{Preparata:1998rz}. In this case, the charge-to-mass ratio could be near to 1. This would produce the most energetic known objects in the universe: the gamma-ray bursts (GRB) with an energy around $10^{54} \text{ ergs }$ ($\simeq 1 M_{\odot} c^2)$ released in few seconds. \us{Beside, we also motivate the investigation of charged collapsing structures from a theoretical perspective as the term due to the charges are interesting in order to constrain the different scenarios of collapse}.
\par
 The outline of the article is the following: in section \ref{sec:generalsolution}, we present the general equations of motion that govern the dynamics of the charged cosmological black hole, they are the generalization of the equations considered for instance in \cite{Firouzjaee:2016hzy}.
 In section \ref{sec:numerics}, we solve numerically the equation of motion for the time evolution of the collapsing structure. In section \ref{sec:special case}, we apply them to special cases of interest to model the cosmological black hole. We report there a new analytic solution and calculate various typical quantities to characterize the properties of the black hole. Our results are applied to a model of gamma ray burst in section \ref{sec:GRB}. In section \ref{sec:ccl}, we sum up our conclusions and propose some perspectives.

\section{General spherically symmetric solution}
\label{sec:generalsolution}
\par
Consider a general inhomogeneous spherically symmetric spacetime \cite{Alfedeel:2009ef} constructed with a charged perfect fluid and a metric expressed in the comoving coordinates, $x^{\mu} = (t,\,r,\,\theta,\,\phi)$:
\begin{equation}\label{metric}
ds^2  = -e^{2\sigma} \, dt^2  + e^{\lambda} \, dr^2 + R^2 \, d\Omega^2 \;,
\end{equation}
 where $\sigma = \sigma(t,r)$, $\lambda = \lambda(t,r)$ are functions to be determined, $R = R(t,r)$ is the physical radius, and $d\Omega^2 = d\theta^2 + \sin^2 \theta \, d\phi^2$ is the metric of the unit 2-sphere.  The energy momentum tensor of the perfect fluid is
 \begin{equation}\label{mT}
   T_M^{\mu\nu} = (\rho + p) \, u^{\mu} \, u^{\nu} + g^{\mu\nu}p \;,
 \end{equation}
 and the electromagnetic tensor is
\begin{equation}
T_{EM}^{\mu \nu} =\frac{1}{4 \pi} \left( F^{\mu \alpha} F^{\nu}\space{}_{ \alpha} -\frac{1}{4}g^{\mu \nu}F_{\alpha \beta} F^{\alpha \beta} \right),
\end{equation}
 where $\rho = \rho(t,r)$ is the mass-energy density, $p = p(t,r)$ is the  pressure, and $u^{\mu} = (e^{-\sigma}, 0, 0, 0)$ is the charged perfect fluid four-velocity. Choosing a perfect fluid implies that there is no heat flow, radiation, or viscosity.
 \par
 The electromagnetic field $F^{\mu \nu} $ satisfies Maxwell's equation\us{s}:
\begin{equation}
 \label{EM1}
\nabla_{\mu} F^{\mu \nu} = 4 \pi J^{\nu}
\end{equation}
and
\begin{equation}
F_{[\alpha \beta, \gamma]}=0,
\end{equation}
where $J^{\nu}$ is the 4-current. To describe the charged black hole, we choose to consider an electric charge at rest in the comoving coordinate\us{s} of \us{the} fluid ; in this case, the potential and the current are given by:
\begin{equation}
\label{eq:q}
A_{\mu}(t,r) = A(t,r) \delta^{0}_{\mu} \hspace{1cm} J^{\nu}=\rho_{EM}(t,r)u^{\nu}.
\end{equation}
\par
The covariant form of the electric field $E_\mu$ is
\begin{equation}
E_\mu=F_{\mu\nu}u^\nu
\end{equation}
\us{From here we assume that the magnetic field is vanishing:
\begin{equation}
B_{\rho} \equiv \frac{1}{2} \epsilon_{\rho \mu \nu \sigma} u^{\mu} F^{\nu \sigma} = 0,
\end{equation}  
}
where $\epsilon^{\mu \nu \alpha \beta}$ is the 4-dimensional totally antisymmetric volume element. We choose the convention $\epsilon_{0123} = \sqrt{- \det g}$ with $\det g$ the determinant of the metric. \us{It is however possible to derive more general equations by using the electromagnetic invariants.}
Along the comoving observer with the fluid, $T_{EM}^{\mu \nu}$ can be written as \cite{ellis} 
\begin{equation}
T_{EM}^{\mu \nu}=\frac12 {\bf E}^2  u^\mu u^\nu+\frac
16  {\bf E}^2  h^{\mu \nu}+\pi^{\mu \nu}.  \label{Tem1}
\end{equation}
$h_{\mu \nu} = g_{\mu \nu} + u_\mu u_\nu$ is the observer hypersurface metric, ${\bf E}^2=E_\mu E^\mu$ is the magnitude of the electric field. $\pi_{\mu \nu}$ is a traceless and space-like symmetric tensor given by
\begin{equation}
\pi^{\mu \nu}_{EM}= \frac{1}{3} {\bf E}^2 h^{\mu \nu}- E^\mu E^\nu.
  \label{Mten1}
\end{equation}
Eq.~(\ref{Tem1}) can be compared with the energy momentum tensor for a generic
imperfect fluid with: 
\begin{eqnarray}
\rho_{EM}(t,r) &=&\frac 1{2} ~ {\bf E}^2 ,  \label{muem}
\\
&&  \nonumber \\
p_{EM}(t,r) &=&\frac 16 ~ {\bf E}^2 ,  \label{pem} \\
&&  \nonumber \\
\pi ^{\mu \nu } &=&\pi^{\mu \nu}_{EM}.  \label{piem}
\end{eqnarray}
\par
Because of the spherical symmetry, the only non-vanishing component of the electromagnetic field is $F^{01}=-F^{10}$ so from Eq.~(\ref{EM1}) we have 
\begin{equation}
F^{01}=e^{-\left(\sigma+\frac{\lambda}{2}\right)} \frac{Q}{R^2}.
\end{equation}
In this article we consider that the only non-vanishing current density is $J^0$, so $Q$ is not an explicit function of time \cite{Bekenstein:1971ej}
\begin{equation}
Q(r)=\int_0^R 4\pi e^{-\left(\sigma+\frac{\lambda}{2}\right)} R^2 J^0 dr.
\end{equation}
%
%

\subsection*{Field Equations} \label{FEq}

The Einstein field equations $G^{\mu\nu} = \kappa \,T^{\mu\nu} - g^{\mu\nu}\Lambda$ can be reduced to the following set of equations:
  \begin{eqnarray}
   e^{2\sigma} \, G^{tt}  &=& - \left( \frac{2 R''}{R } + \frac{R'^2}{R^2} - \frac{R'}{R} \, \lambda' \right) \, e^{-\lambda} \nonumber{} \\&& + \left( \frac{\dot{R}^2}{R^2} + \frac{\dot{R}}{R} \, \dot{\lambda} \right)
      \, e^{-2\sigma}  
    + \frac{1}{R^2} = \kappa \rho +\frac{Q^2}{R^4} + \Lambda ~,\nonumber{} \\
      \label{tt} 
            \end{eqnarray}
            
             \begin{equation}
   e^{\lambda} \, G^{tr}  = \left( \frac{2 \dot{R}'}{R} - \frac{2 \dot{R}}{R} \, \sigma'
      - \frac{R'}{R} \, \dot{\lambda} \right) \, e^{-2\sigma} = 0 ~,
      \label{zero} \\
        \end{equation}
 \begin{eqnarray}
   e^{\lambda} \, G^{rr}  &=& \left( \frac{R'^2}{R^2} + \frac{2 R'}{R} \, \sigma' \right) \, e^{-\lambda} \nonumber{}\\&& - \left( \frac{2 \ddot{R}}{R} + \frac{\dot{R}^2}{R^2} - \frac{2 \dot{R}}{R} \, \dot{\sigma} \right)
      \, e^{-2\sigma}
     - \frac{1}{R^2} = \kappa p -\frac{Q^2}{R^4}- \Lambda ~, \nonumber{}\\
      \label{rr} 
       \end{eqnarray}
 \begin{eqnarray}
   R^2 \, G^{\theta\theta}  = \left( \frac{R''}{R} + \frac{R'}{R} \, \sigma' + \sigma'' + \sigma'^2
      - \frac{R'}{2 R} \, \lambda'  - \frac{1}{2} \, \sigma' \, \lambda' \right) \, e^{-\lambda}
      \nonumber \\
   ~~~ + \left( \frac{\dot{R}}{R} \, \dot{\sigma} - \frac{\ddot{R}}{R} - \frac{1}{2} \, \ddot{\lambda}
          + \frac{1}{2} \dot{\lambda} \, \dot{\sigma} - \frac{\dot{R}}{2 R} \dot{\lambda}
      - \frac{1}{4} \, \dot{\lambda}^2 \right) \, e^{-2\sigma} = \nonumber \\
     \kappa p +\frac{Q^2}{R^4} - \Lambda ~.\nonumber\\
      \label{theta}
 \end{eqnarray}
 
The conservation equations are
    \begin{equation}
   \frac{2 e^{2\sigma}}{(\rho + p)} \, \nabla_\mu T^{t\mu}  = \dot{\lambda}
      + \frac{2 \dot{\rho}}{(\rho + p)} + \frac{4 \dot{R}}{R}\; = 0,
      \label{gtt}
       \end{equation}
   \begin{equation}
   \frac{e^{\lambda}}{(\rho + p)} \, \nabla_\mu T^{r\mu}  = \sigma' + \frac{p'}{p + \rho}-\frac{QQ'}{4\pi (\rho + p)R^4} = 0,
      \label{grr}
 \end{equation}
where the dot denotes a partial derivative with respect to $t$, and the prime denotes a partial derivative with respect to $r$. Using $\dot{Q}=0$, Eq.~(\ref{rr}) leads to the following equation
\begin{equation}\label{Pr}
   \frac{\partial}{\partial t}\left[ R + R \dot{R}^2 e^{-2\sigma}-RR'^2 e^{-\lambda}+\frac{Q^2}{R}-\frac{1}{3}\Lambda R^3\right] = -\kappa p R^2\dot{ R}. \;
   \end{equation}
The term in the brackets is related to the Misner-Sharp mass $M$ \cite{Bekenstein:1971ej}
 \begin{equation}
   \frac{2M}{R} = \dot{R}^2 e^{-2\sigma} - R'^2 e^{-\lambda} + 1 +\frac{Q^2}{R^2}- \frac{1}{3} \Lambda R^2 \;.
   \label{mR}
 \end{equation}
 Eq.~(\ref{Pr}) can be written as
 \begin{equation}
   \kappa  p  = -\frac{2\dot{M}}{R^2 \dot{R}} ~,   \label{pressure}
 \end{equation}
and Eq.~(\ref{tt}) can be written as follows
  \begin{equation}
   \kappa  \rho+ \frac{QQ'}{R^3}  = \frac{2M'}{R^2R'} ~.   \label{denn}
 \end{equation}
 
 When $\Lambda=0$ and $R(t,r)=r$,  Eq.~(\ref{mR}) reduces to the familiar Reissner-Nordstr$\ddot{\text{o}}$m solution
  \begin{equation}
   e^{\lambda}= \frac{1}{1-\frac{2M}{r}+\frac{Q^2}{r^2} } \;.
   \label{RNm}
 \end{equation}
 
\par
After simplification of the conservation laws and Einstein equations, the five coupled partial differential equations governing the evolution of the CCBH are:
  \begin{equation}
   \label{Ev1}
      \dot{R} = \pm e^{\sigma} \sqrt{\frac{2M}{R} -\frac{Q^2}{R^2}+ 2E+ \frac{\Lambda R^2}{3}} ~, \\
 \end{equation}
 \begin{equation}
   \dot{M} = \frac{- \kappa p \, \dot{R} R^2}{2} ~,
 \end{equation}
 \begin{equation}
   \dot{\rho} = - p' \, \frac{\dot{R}}{R'}+\frac{2\dot{R}QQ'}{4\pi R^4}
      - (\rho + p) \left[ \frac{\dot{R}'}{R'} + \frac{2 \dot{R}}{R} \right] ~,
      \label{rhodot}
 \end{equation}
 \begin{equation}
   \dot{p} = \frac{dp}{d\rho} \, \dot{\rho}  ~,
 \end{equation}
 \begin{equation}
   \dot{\lambda} = \frac{2}{R'} \left( \frac{ p' \dot{R}}{(\rho + p)} + \dot{R}' -\frac{2\dot{R}QQ'}{4\pi (\rho + p) R^4}\right)~,
 \end{equation}
  where
\begin{equation}
   \label{f}
      2E(t,r) = R'^2 e^{-\lambda} - 1~
 \end{equation}
 is the curvature term, analogous to $E(r)$ in the LTB model. It is not to be confused with the electric field of section \ref{sec:generalsolution} also written $\textbf{E}$ but always bolded. According to (\ref{f}) and the LTB coordinate conditions, the choice of
 $\lambda_{0}(r)$ is equivalent to the choice of $E(t_0,r)=E_0(r)$. Moreover, $\sigma(t,r)$ is calculated from Eq.~(\ref{grr}) by integrating along constant $t$ with $\sigma(t,r_0)=0$ \cite{Moradi:2015caa}. Note that 4 initial functions need to be determined, $R_0(r)$, $\rho_0(r)$, $\lambda_0(r)$ and $\sigma_0(t)$, as well as the equation of state $p(\rho)$ and $Q(r)$.\\
 
\par
 In the study of the accretion of a two-component fluid into a compact object, when some small amplitude pulsation of the fluid component exists, considering the solutions with $Q =Q(t,r)$ would be very important because the presence of a pulsation is equivalent to have $\dot{Q} \neq 0 $ \cite{Ludwig:2014una}.

 \section{Numerical results}
\label{sec:numerics}
To run our numerical code, we first discuss here the equation of state of the model which is the tool to handle the pressure behavior in relation with the energy density.
However, we do not study the effect of different equations of states on the rate of collapse and black hole evolution, in the case of charged black hole, it is left for further studies. In the case of neutral black hole, the comprehensive argument for the pressure behavior can be found in \cite{Moradi:2015caa, Moradi:2013gf}. Here we consider the following equation of state:
\begin{equation}
\label{eq:eos}
p(t,r)=\omega f(r) \rho(t,r),
\end{equation}
 with $\omega=\frac{1}{10}$ and $f(r)$ vanishing in the FLRW limit $f(r\to \infty)=0$: the pressure is zero at infinity as we consider a dust source for the expansion of the universe.
The Hubble parameter that corresponds near the black hole to the collapse rate, is defined as $H(t,r)=\frac{\dot{R}(t,r)}{R(t,r)}$.

\subsection{Initial conditions}

In order to fulfill the FLRW limit and to have a structure with a void, we choose the the initial density as follows 
\begin{equation}
\label{eq:initdp}
\rho(t_0,r)= \rho_c + \rho_s -r^2\rho_G(r),
\end{equation}
where  $\rho_c$ is the background density, $\rho_s$ is the density of the collapsing object and $\rho_G(r)=a \exp(-r^2)$ is a Gaussian term that controls the location of the void. $a$ is a dimensionless normalization constant. 
 The initial conditions for the curvature term $E(t,r)$ and for the physical radius $R(t,r)$ are:
\begin{eqnarray}
&E(t_0,r)=- b_0r^2 e^{-b_1 r}\nonumber\\
&R(t_0,r)=r,
\end{eqnarray}
$b_0$ and $b_1$ are constants. Observe again that the FLRW limit is fullfiled for $r \rightarrow \infty$. We now turn to the numerical resolution of the differential system derived in section section \ref{FEq}. 
\subsection{Solving the equations of motion (\ref{Ev1})-(\ref{f})}
\begin{figure}
\typeout{*** EPS figure 1}
\begin{center}
\includegraphics[scale=0.5]{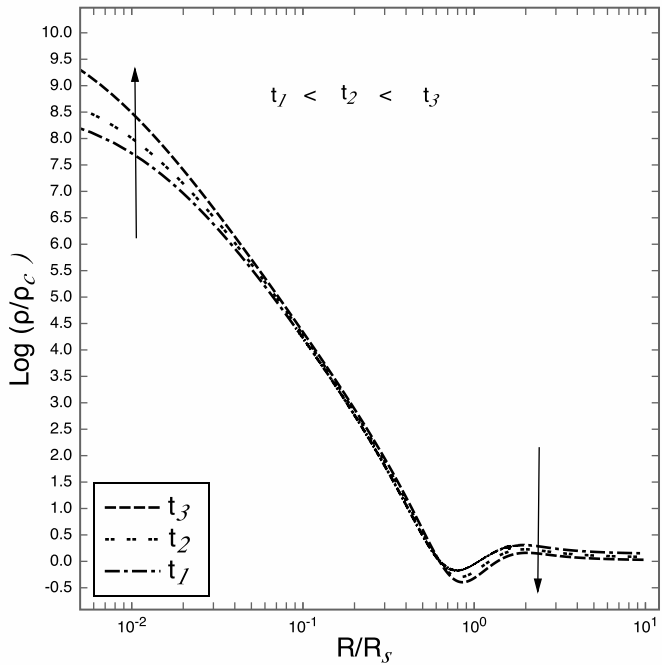}
\caption{Evolution of the density of the CCBH \us{with pressure} with an initial density profile given by (\ref{eq:initdp}) with $\rho_s=\frac{\rho_c}{r^2}$, $a=\rho_c$. In the FLRW limit, the density is decreasing while it is increasing inside the structure. $R_s$ is the radius of the collapsing structure, $\rho_c$ is the background density.  For decreasing $R(t,r)$, one can see first the decreasing FLRW density corresponding to the expanding universe, second a void and third an increasing density for the gravitational collapse. The equation of state is given by equation (\ref{eq:eos}) with $f(r)=\exp(-r)$. $\xi=\frac{Q_{BH}}{M_{BH}}=0.1$}
\label{fig:3}
\end{center}
\end{figure}

To solve the equations under consideration, we modified a code developed in \cite{Moradi:2013gf}. Figure \ref{fig:3} shows a typical time evolution of the energy density. The two important features to be noted are a decrease of the energy density for large $R$ and an increase for smaller $R$, this illustrates the collapse process in an otherwise expanding universe. 

The \us{numerical investigations} of these solutions show\us{ed} that when the pressure is zero and the charge is \us{large compared} to the mass of the collapsing structure, the energy density behaves in the same way that \us{in} the case of a small pressure and \us{large} charge. Therefore, we conclude that the presence of a small pressure does \us{not} significantly affect the system under study. While we presented the pressure terms in section \ref{sec:generalsolution} for completeness, we will from now on assume $p=0$. It will allow us to derive analytic solutions in the next section. Mathematically speaking, the absence of pressure can be understood in the following way: from the definition of the Misner-Sharp mass Eq.~(\ref{mR}), $M \propto Q^2$. Since we assumed $\dot{Q} = 0$, $Q$ cannot contribute in the pressure part (\us{see} Eq.~(\ref{pressure})), hence it does \us{not} contribute in a term like matter pressure. We however note that, if one changes the definition of \us{the} mass, for instance by considering the Hawking mass \cite{Haw}, the electromagnetic pressure behaves exactly like matter pressure, hence changing the above statement.
\section{Solutions with zero pressure}
\label{sec:special case}
In section \ref{sec:generalsolution}, we presented the Einstein equations with a general charged fluid in the spherically symmetric case. We then solved numerically the equations in section \ref{sec:numerics}. \us{The study of those numerical results motivated us to investigate the pressureless case which allows for analytic solutions.} In this section we \us{therefore} consider two famous special cases: a dust charged black hole and a point mass charged black hole in de Sitter spacetime. We checked that the numerics corresponds to those analytic cases.
\subsection{Charged LTB metric}

\label{sec:chargeLTBm}
\paragraph*{}
Assuming $\Lambda=0$, $p=0$ and $Q=\text{const}$ the metric (\ref{metric}) reduces to the charged LTB metric: 

\begin{equation}
ds^2=-dt^2+\frac{R'^2(t,r)}{1+2E(r)}dr^2+R^2(t,r) d\Omega^2\us{,}
\end{equation}
\us{which is discussed in \cite{Pleb}, see in particular p.~374.} The signature (- +++) of the metric implies that $\forall r, E(r) > -\frac{1}{2}$. The key assumption here is that the electromagnetic energy is negligible relative to the dust fluid. The general case for the electromagnetic field is described with a fluid with pressure, a heat flow and an anisotropic pressure \cite{ellis}. We note also that another more general treatment for the charged LTB metric could include magnetic monopoles however they are not relevant for the purpose of this paper.
Furthermore, the assumption $Q=\text{const}$ implies that the universe is charged in contradiction with the standard cosmological model. We stress here that the model we consider is a toy model to understand the basic properties of a charged cosmological black hole but that a realistic universe would be filled by numerous black holes which charges would screen each others. In other words, for our purpose, we consider the idealized case $Q=\text{const}$, but the effective $Q$ in the universe in more realistic models would be zero at infinity.

 For our purpose of describing charged black holes, the system to solve becomes:
 
\begin{align}
\label{Rdotltb}
& \dot{R}^2(t,r)=2E(r) +  \frac{2 M(r)}{R} - \frac{Q^2}{R^2(t,r)} , \\
& M'(r)= \frac{1}{2}\kappa \rho R^2(t,r) R'(t,r),
\end{align}
since $\dot{M}=0$, $M$ is only a function of $r$, \textit{i.e.}~$M=M(r)$.
\par
The explicit solutions of Eq.~(\ref{Rdotltb}) involve elliptic function which in the case of $Q=0$ were discussed by Lemaitre \citep{Lemaitre:1933gd} and Omer \citep{Omer}. When $Q\neq 0$ we obtain the explicit solutions as follow

\begin{itemize}

\item $E(r) < 0$:

\begin{equation}
\label{equ:solnew}
 \left\{
  \begin{array}{@{}ll@{}}
  R(t,r) = \frac{M(r)}{2 E(r)} \left(\cos \eta -1+ \frac{ E(r) e^{i \eta} Q^2 }{  M(r)^2 }\right) \\
    \eta - \sin \eta  + \frac{i E(r) e^{i \eta} Q^2 }{  M(r)^2 }= \frac{(-2 E(r))^{3/2}}{M} \left[t-t_B (r)\right].
  \end{array}\right.
\end{equation}

\item $E(r) = 0$:
\begin{eqnarray}
R(t,r) = \frac{1}{6} \left[\frac{5 Q^2}{M(r)}+\frac{Q^4}{ M(r)^2 L(t,r)}+L(t,r)\right], \nonumber{} \\
L(t,r) = \nonumber{} \\
\text{\tiny $
\left(486 \left[t-t_B(r)\right]^2 M(r)-\frac{Q^6}{M(r)^3}- \frac{18 \sqrt{3} \sqrt{243\left[t-t_B(r)\right]^4 M(r)^4 -\left[t-t_B(r)\right]^2 Q^6}}{M(r)}\right)^{1/3}.$}  \nonumber{} &&
\end{eqnarray}

\begin{eqnarray}
\end{eqnarray}

\item $E(r) > 0$:

\begin{equation}
\label{eq:solnewf}
 \left\{
  \begin{array}{@{}ll@{}}
R(t,r) = \frac{M(r)}{2 E(r)} \left(\cosh \eta -1+ \frac{ E(r) e^{-\eta} Q^2 }{ M(r)^2 }\right) \\
   \eta - \sinh \eta +  \frac{ E(r) e^{-\eta} Q^2 }{M(r)^2 } = -\frac{(2 E(r))^{3/2}}{M(r)} \left[t-t_B (r)\right].
  \end{array}\right.
\end{equation}

\end{itemize}

\us{The results of equations (\ref{equ:solnew})-(\ref{eq:solnewf}) are the main results of this article. They represent new solutions of the charged LTB metric.}
Clearly if $Q=0$ these solutions reduce to the LTB solutions \cite{Pleb}. 
\subsubsection{Characterization of the horizons}
\label{sec:Boundary}
Here we study in more details the horizon of the CCBH. We will not review the whole theory of evolving black hole horizons here. A comprehensive discussion can be found in \cite{Firouzjaee:2011dn} and the references therein. The expansion for  ingoing and outgoing null geodesics is:
$\theta_{(\ell)}\propto \left(1-\frac{\sqrt{\frac{2M}{R}+2E-\frac{Q^2}{R^2}}}{\sqrt{1+2E}}\right)$,
$\theta_{(n)}\propto \left(-1-\frac{\sqrt{\frac{2M}{R}+2E-\frac{Q^2}{R^2}}}{\sqrt{1+2E}}\right)<0$.
The sign of $\theta_{(\ell)}$ is the same as the one of the quadratic polynom in $R$: $1=\frac{2M}{R}-\frac{Q^2}{R^2}$. Its roots are 
\begin{equation}
\label{hor}
R_\pm=M\pm \sqrt{M^2-Q^2}. 
\end{equation}
To study the horizon of the CCBH, we consider $R =R_+$ where the expansion for null outgoing
geodesic changes its sign.
Furthermore, the ingoing null geodesics expansion is negative everywhere. Therefore, the 3-manifold $R = R_+$ is a \emph{marginally trapped tube} (MTT).
\par
\begin{figure}
\typeout{*** EPS figure 2}
\begin{center}
\includegraphics[scale=1]{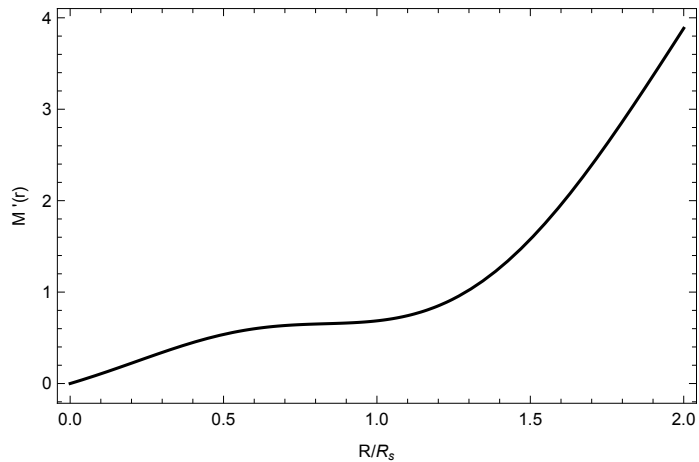}
\caption{A typical behavior of $M'(r)$ for a \us{pressureless CCBH} . We see that $M'(r)$ increases when $R$ is increased. For compact objects $M'(r)$ is always increasing: this qualitative behavior is not model dependent.}
\label{fig:5}
\end{center}
\end{figure}

 
Now we prove that, the MTT is located between the singularity line $R=0$ and the boundary between the collapsing and the expanding region, namely $\dot{R}=0$. Imposing $\dot{R}=0$ in Eq.~(\ref{Rdotltb}), it is sufficient to have $2E>-1$ (that is no changes of the metric's signature) to obtain the apparent horizon, $1=\frac{2M}{R}-\frac{Q^2}{R^2}$ as discussed before. This horizon is space-like asymptotically tending to be light-like at late times when the matter flux decreases. This can best be seen by comparing the slope of the apparent horizon relative to the light cone at every coordinate point of it. This result is in contrast with the Schwarzschild black hole horizon where the apparent horizon is always light-like: a null surface.
At late times, however, we expect the apparent horizon to become approximately light-like and approaching the event horizon. Mathematically, we calculate the apparent horizon by considering $\frac{dt}{dr}|_{AH}=-\frac{R'-\left(M'+\frac{M M'}{\sqrt{M^2-Q^2}}\right)}{\dot{R}-\left(\dot{M}+\frac{M \dot{M}}{\sqrt{M^2-Q^2}}\right)}.$ From Eq.~(\ref{Rdotltb}) at horizon: $\dot{R}|_{AH}=-\sqrt{1+2E(r)}$, we find 
\begin{equation}
\frac{dt}{dr}|_{AH}=\frac{R'-\left(M'+\frac{M M'}{\sqrt{M^2-Q^2}}\right)}{\sqrt{1+2E(r)}},
\label{dtdr1}
\end{equation}
 We now discuss the nature of the apparent horizon, and to do so, the behavior of all the quantities involved in equation (\ref{dtdr1}) is required. They are all straightforward but $M'(r)$ so we will use again the numerical code of section \ref{sec:numerics} to characterize it. We propose a typical behavior of $M'(r)$ for a CCBH in figure \ref{fig:5}.

 Since $M' > 0$,  $\frac{dt}{dr}|_{AH}=\frac{R'-\left(M'+\frac{M M'}{\sqrt{M^2-Q^2}}\right)}{\sqrt{1+2E(r)}} < \frac{dt}{dr}|_{null}=\frac{R'}{\sqrt{1+2E(r)}}$ at all times. From Eq.~(\ref{hor}) when $Q$ is increased, the radius of the black hole $R$ decreases, hence according to Fig.~(\ref{fig:5}) $M'$ decreases for decreasing $R$, so the value of $R'-\left(M'+\frac{M M'}{\sqrt{M^2-Q^2}}\right)$ tends to $R'$. So, by increasing the value of $Q$ the black hole horizon tends to an isolated horizon in a shorter time comparing to the pure dust case. Therefore, the apparent horizon is always a space-like dynamical horizon leading to a slowly varying horizon at late times \cite{Booth:2003ji}.

\us{
\subsubsection{Study of the singularity}
}
\par 
\us{Now, we discuss the nature of the singularity for the charged LTB case. We consider a falling observer, its equation of motion is given by equation (\ref{Ev1}). Inspecting it, one can see that after passing the inner and outer horizon, for decreasing $R(r,t)$, the observer reaches a turning point: $\dot{R}(t,r)=0$ and its motion that was initially inward becomes onward. The observer, as in the RN case eventually goes out of the black hole into another external universe, see \cite{poisson} for more details in the case of RN. As the falling observer is comoving with respect to the perfect fluid, the same reasoning apply in the case under study in this paper. 
}
\par \us{ The singularity occurs at $R=0$ but is never reached by the observer. The tangent vector to the singularity is
\begin{align}
dt/dr|_{sin}=\us{-}\frac{R'}{\dot{R}}.
\end{align}
Beside, the null geodesics tangent vector is
\begin{align}
dt/dr|_{null}=\frac{R'}{\sqrt{1+2E}}.
\end{align}
A comparison between the two tangent vectors gives
\begin{align}
\frac{dt/dr|_{sin}}{dt/dr|_{null}}	= -\frac{\sqrt{1+2E}}{\dot{R}}= -\frac{\sqrt{1+2E}}{\sqrt{2E + \frac{2M(r)}{R} - \frac{Q^2}{R^2}}}.
\end{align}
From the discussion of the in-going geodesic of section \ref{sec:Boundary}, one can conclude that $\left|\frac{dt/dr|_{sin}}{dt/dr|_{null}}\right| <1$, therefore black hole singularity is timelike for any value of the charge $Q$.
}

\subsubsection{Matter flux on the apparent horizon}
To observe the effect of the charge on the black hole accretion we calculate the matter flux on the apparent horizon, 

\begin{equation}
\label{matterflux}
\frac{dM}{dt}|_{AH} = \frac{\partial M}{\partial t}|_{AH}  + \frac{\partial M}{\partial r}|_{AH} \frac{dr}{dt}|_{AH} =  M'|_{AH} \frac{dr}{dt}|_{AH}.
\end{equation}
As we discussed before, for big $Q$ at horizon $M'$ decreases. Moreover, $\frac{dt}{dr}|_{AH}$ increases faster for larger $Q$ so $\frac{dr}{dt}|_{AH}$ decreases as $Q$ increases. Therefore we conclude that the bigger the charge $Q$ the smaller the matter flux at horizon.
\subsubsection{Calculation of the C-function}
To quantify the growth of the apparent horizon, we consider the future-directed 
outgoing and ingoing null vectors, $\ell^a$ and  $n^a$ respectively, and
	the expansions $\theta_{\ell}$ and $\theta_n$  of the null curves generated by 
	these vectors which are cross-normalized, $ \ell.n=-1 $. See \cite{Firouzjaee:2011dn} and references therein for more details.
	\par
\begin{figure}
\typeout{*** EPS figure 3}
\begin{center}
\includegraphics[scale=1]{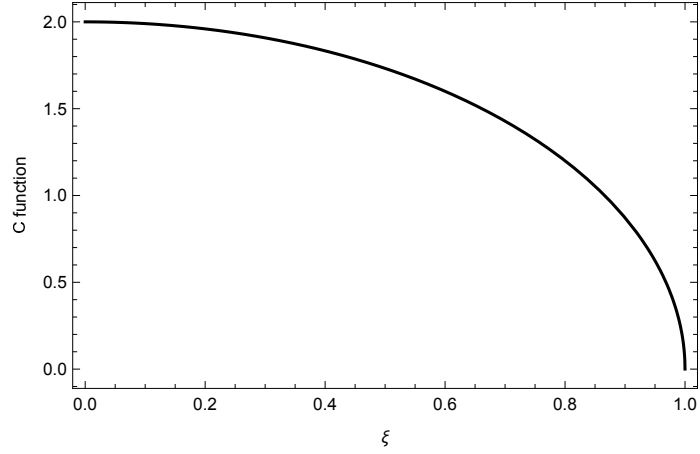}
\caption{$C$ function of a \us{pressureless CCBH} for different values of $\xi=\frac{Q_{BH}}{M_{BH}}$. From this behavior \us{of} the $C$-function, one can conclude that the charge helps to reach the event (isolated) horizon in a shorter time.}
\label{fig:4}
\end{center}
\end{figure}

	
	Let $V^a$ be tangential to the apparent horizon, and orthogonal to the foliation by 
	marginally trapped surfaces. It is always possible to find a function $C$ such that $V^a = \ell^a - C n^a$. Moreover, 
	the definition of $V^a$ leads to $\pounds_V \theta_\ell = 0$, which gives an expression for $C$:
\begin{equation} 
	C = \frac{\pounds_\ell \theta_\ell}{\pounds_n \theta_\ell} . 
	\label{defC}
\end{equation}
When $C<0$ the apparent horizon is an inner apparent horizon, when $C>0$ the apparent horizon is an outer apparent horizon, and when $C=0$ it becomes an event (isolated) horizon \cite{Firouzjaee:2014zfa}. The value for the dimensionless $C$ function is important because it shows the type of the black hole horizon. \us {This $C$ function of this work corresponds to the $\frac{C}{B}$ quantity of Ref.~\cite{Nielsen:2005af}. It becomes dimensionless and shows the evolving horizon's properties}. Using the Raychaudhuri equation (\ref{defC})
\begin{equation}
\label{cf}
C=\dfrac{T_{ab}\ell^a \ell^b}{1/2A-T_{ab}\ell^a n^b}|_{AH}=\dfrac{\rho}{1/8 \pi R^2-\dfrac{\rho}{2}}|_{AH} =\dfrac{2M'}{R'-M'}|_{AH},
\end{equation}
where $A=4 
\pi R_+^2$ is the area of the black hole.  For $Q=const$, Eq.~(\ref{cf}) is
 \begin{equation}
\label{cf1}
C= 2 \sqrt{1-\frac{Q^2}{M^2}} \Bigg \lvert_{AH}.
\end{equation}
Clearly increasing the value of $Q$ decreases $C$.  Therefore, one can see that charge helps to have less flux relative to the pure dust case, in this sense the charge can even screen the black hole formation. A graphical representation of the $C$-function can be found in figure \ref{fig:4}.

This concludes our study of the charged LTB metric in connection with CCBH. We now turn to another popular model: we assume that the background instead of being FLRW is simply a de Sitter space.
\subsection{CCBH in de Sitter background}
\label{sec:dSBH}
A cosmological constant is the simplest model to describe the current cosmic
acceleration which has been observed first from supernovae and then confirmed by various cosmic observables. As discussed in section \ref{sec:intro}, many physical processes usually taken in flat Minkowski spacetime are now investigated in de Sitter space. The de Sitter metric is an exact vacuum solution to the Einstein equations with a cosmological constant. We will consider it as a special case of the FLRW metric, indeed as the universe is currently expanding at an accelerated rate, it will reach asymptotically a de Sitter metric. 
\par
The Reissner-Nordstr$\ddot{\text{o}}$m solution can be extended to the de Sitter static solution background as a de Sitter-Reissner-Nordstr$\ddot{\text{o}}$m solution. The metric in static coordinates is 
\begin{equation}
\label{a3}
ds^{2}  = -\Phi dt^{2} + \Phi^{-1} dR^{2} + R^{2} d \Omega^{2} ,
\end{equation}
where 
\begin{equation}
\label{ }
\Phi =   1- \frac{\Lambda}{3} R^{2} - \frac{2M}{R} +\frac{Q^{2}}{R^{2}},
\end{equation}
where $\Lambda$ is cosmological constant. This metric does \textit{not} describe a charged black hole in an otherwise expanding universe. \us{In \cite{Firouzjaee:2016hzy}, the coordinate transformation in order to obtain a point mass CCBH in a cosmological background was presented in the case of charged and uncharged BH solutions. We present here only the final result for the metric:}
\begin{widetext}
\label{eq:BHdS}
$ds^{2}  = - d \tau ^{2} +   \text{\tiny $ \left(\frac{\Lambda}{3}   \frac{e^{-2 \left(r + \tau\right)\sqrt{\frac{\Lambda}{3}}}  \left[e^{4 \left(r + \tau\right) \sqrt{\frac{\Lambda}{3}}} +  3  \Lambda Q^{2}\right]  }{2 \Lambda}      -  \frac{2 \Lambda Q^{2}}{e^{-2 \left(r + \tau\right)\sqrt{\frac{\Lambda}{3}}}  
		\left[e^{4 \left(r + \tau\right)\sqrt{\frac{\Lambda}{3}}} +  3  \Lambda Q^{2}\right]   } \right)d r^{2} $} + \left( \frac{e^{-2 \left(r + \tau\right)\sqrt{\frac{\Lambda}{3}}}  \left[e^{4 \left(r + \tau\right)\sqrt{\frac{\Lambda}{3}}} +  3  \Lambda Q^{2}\right]  }{2 \Lambda} \right) d \Omega^{2}.$
\end{widetext}
This metric is a point mass case of a CCBH in the expanding de Sitter background. The general dynamic models of CCBH of section \ref{sec:generalsolution} reduce to (\ref{eq:BHdS}) in the de Sitter case therefore studying its properties could give much insight for general properties of the CCBH. For instance a good starting point to study thermodynamic properties of the CCBH would be with the use of this de Sitter limit. More properties and motivation to investigate this metric are presented in Ref.~\cite{Firouzjaee:2016hzy}.
\section{Charged black hole and Gamma-Ray Bursts}
\label{sec:GRB}
The fireshell model is an alternative model introduced to propose an engine for Gamma-Ray Bursts (GRB) through the induced gravitational collapse of compact objects like neutron stars leading to the formation of a Kerr-Newman BH \cite{Ruffini:2016jmq}. In this model, the energy released in the GRB comes from Schwinger pair production \cite{Schwinger} around a charged black hole. We have motivated the existence of charged black holes in the introduction within the paradigm of the induced gravitational collapse which is based on the fireshell model. Once the electromagnetic field becomes overcritical,  a substantial number of pairs will be created in the region around the BH \cite{Preparata:1998rz, Ruffini:2009hg} (see also references therein), called ``dyadosphere'' in the case of a Reissner-Nordstr$\ddot{\text{o}}$m BH.
\par
With the solution for $R(t, r)$ discussed in section \ref{sec:numerics} and \ref{sec:special case}, it is then possible to define a dyadosphere for the CBH. The dyadosphere is defined as \us{the} region between \us{the} outer horizon and the largest sphere where the electrical field is overcritical around the BH \emph{i.e.}~the radius which is such that:
\begin{equation}
\Big|{\bf E}[R(t,r)]\Big| \geq {\bf E_c} \equiv \frac{m^2_e}{e} \simeq 1.268 \times 10^{-9} \text{ m}^{-1},
\end{equation}
in geometric units.
 In modeling a CCBH we did not consider the rotation of the BH, so after forming an isolated horizon, one expects a dyadosphere with a radius of 

\begin{equation}
R_{ds} \simeq 1.12 \times 10^6 \sqrt{\mu \xi} \text{ m},
\end{equation}
with $\mu = \frac{M_{BH}}{M_{\odot}}$ and $\xi = \frac{Q_{BH}}{M_{BH}}$
\par
\begin{figure}
\typeout{*** EPS figure 4}
\begin{center}
\includegraphics[scale=.5]{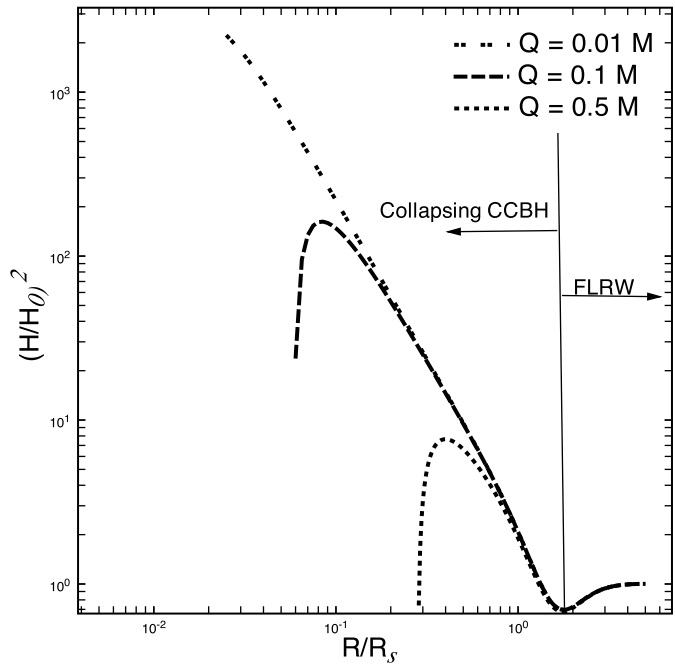}
\caption{The effect of the electric charge ($Q$) on the collapse rate ($H=\frac{\dot{R}}{R}$)  \us{of a CCBH with pressure} at a constant time for three different values of $Q$. When $Q$ is large, the collapse rate tends to zero for large radius $R$. This illustrates the repulsive nature of the electromagnetic energy. In the case $Q=M$, this repulsive force halts the collapse and the BH does not form. Note also that the equation of state for \us{the} three cases is the same.}
\label{fig:2}
\end{center}
\end{figure}

Now we apply the numerical investigation of section \ref{sec:numerics} to define the dyadosphere. As seen from figure \ref{fig:2} when $Q$ is large enough to produce the overcritical electric field required for vacuum polarization to happen, we report a strong repulsive electric force, an explosion, preventing any collapse and formation of a BH at all. Therefore within the toy model developed in this article, it is not possible to constrain the induced gravitational collapse. We are in this view in perfect agreement with the classical work of Wald: the presence of a large charge prevents once again the gravitational collapse \cite{Wald:1974np}. We however hope to come back to those issues in a future paper, it could be possible, in order to better model the collapse to consider the pressure together with the charge distribution. If such approximations are still inconclusive to obtain a GRB, it is eventually the full Einstein-Maxwell system which will have to be solved numerically without any symmetry assumption contrary to the work in this article.

\section{Conclusions and perspectives}
\label{sec:ccl}
The cosmological black hole is a topical issue nowadays as it can reveal dynamical features absent in classical (astrophysical) black hole theory. They are therefore particularly relevant for the study of a gravitational collapse and for the formation of black holes. In this paper we presented the charged black holes which is a natural step toward studying cosmological black holes located in an expanding cosmological background. As the cosmological background evolves with time, the dynamics is a characteristics of these black holes.
\par
We first presented the general equations of motion for the charged cosmological black hole. Second we solved them numerically in a specific case which can be found in figure \ref{fig:3}. Qualitatively, it was possible to observe the collapse and a decrease of the density way outside the black hole. Third we presented a new analytic solution to the charged LTB equations\us{, they can be found in equation (\ref{equ:solnew})-(\ref{eq:solnewf})}. This solution can describe a CCBH. Since the neutral matter moves on geodesics different than the ones for a pure gravitational field (without charge), this is a pure relativistic effect of the charge. In order to investigate the effect of the charge on the CBH, we computed various quantity such as the density evolution, the flux of these black holes and the C-function. We also characterized the effect of the charge on the type of singularity and of horizon. The main results are that (i) the singularity is always time-like, (ii) the horizon is space-like asymptotically reaching a slowly varying horizon and (iii) the presence of a charge decreases the matter flux of the black hole. We also presented the special case of a CCBH in a de Sitter background space-time which corresponds to the asymptotic state of the background.
\par
Forth and last, we applied this model of CCBH to an idea to model GRB: the fireshell model, we do not report any possibility to trigger the events predicted by the fireshell model within our working hypotheses. We however suggest that it could be possible to further challenge this model by considering more general solutions with non zero pressure and/or more involved charge distribution. It could be also possible to model more precisely the pair creation process within the collapse in order to avoid the halt of it too soon to release a GRB. Those consideration are left for future works.
\par 
Other routes opened by this article includes the exploitation of the novel solution found in equation (\ref{equ:solnew})-(\ref{eq:solnewf}) which could be relevant not only in black hole modeling or in cosmology but also for nuclear physicists where the LTB solution is also used.  Regarding cosmology, the LTB metric is sometimes used to model structures such as void or overdensities. This solution offers now the possibility to model charged structure for instance hydrogen clouds and work out the effect of their charge to cosmological observables. While for this work, we imposed a specific charge distribution (see equation (\ref{eq:q})), it is desirable to study the general case for the formation of a charged black hole via numerical simulations of the full Einstein-Maxwell system. \\

\acknowledgments
CS thanks the Institut d'Astrophysique Spatiale at Paris for its hospitality and Maxime Enderli for discussions about GRBs.

\end{document}